\documentclass[twocolumn, 3p]{elsarticle}
\usepackage{color}
\usepackage{amsfonts}
\usepackage{amsmath}
\usepackage{epstopdf}
\begin{document}


\newcommand{\beq}{\begin{equation}}
\newcommand{\eeq}{\end{equation}}
\newcommand{\beqa}{\begin{eqnarray}}
\newcommand{\eeqa}{\end{eqnarray}}
\newcommand{\lf}{\hfil \break \break}
\newcommand{\ahat}{\hat{a}}
\newcommand{\adag}{\hat{a}^{\dagger}}
\newcommand{\adagg}{\hat{a}_g^{\dagger}}
\newcommand{\bhat}{\hat{b}}
\newcommand{\bdag}{\hat{b}^{\dagger}}
\newcommand{\bdagg}{\hat{b}_g^{\dagger}}
\newcommand{\chat}{\hat{c}}
\newcommand{\cdag}{\hat{c}^{\dagger}}
\newcommand{\nhat}{\hat{n}}
\newcommand{\xhat}{\hat{x}}
\newcommand{\phat}{\hat{p}}
\newcommand{\Pihat}{\hat{\Pi}}
\newcommand{\rhohat}{\hat{\rho}}
\newcommand{\lambdahat}{\hat{\lambda}}
\newcommand{\shat}{\hat{\sigma}}
\newcommand{\ket}[1]{\mbox{$|#1\rangle$}}
\newcommand{\bra}[1]{\mbox{$\langle#1|$}}
\newcommand{\ketbra}[2]{\mbox{$|#1\rangle \langle#2|$}}
\newcommand{\braket}[2]{\mbox{$\langle#1|#2\rangle$}}
\newcommand{\bracket}[3]{\mbox{$\langle#1|#2|#3\rangle$}}
\newcommand{\mat}[1]{\overline{\overline{#1}}}
\newcommand{\hak}[1]{\left[ #1 \right]}
\newcommand{\vin}[1]{\langle #1 \rangle}
\newcommand{\abs}[1]{\left| #1 \right|}
\newcommand{\tes}[1]{\left( #1 \right)}
\newcommand{\braces}[1]{\left\{ #1 \right\}}
\newcommand{\sub}[1]{{\mbox{\scriptsize #1}}}
\newcommand{\com}[1]{\textcolor{red}{[\textit{#1}]}}
\newcommand{\novel}[1]{\textcolor{blue}{#1}}
 
\begin{frontmatter}


\title{Assessments of macroscopicity for quantum optical states}

\author[dtuaddress]{Amine Laghaout\corref{mycorrespondingauthor}}
\author[dtuaddress]{Jonas S. Neergaard-Nielsen}
\author[dtuaddress]{Ulrik L. Andersen}
\address[dtuaddress]{Department of Physics, Technical University of Denmark, Building 309, 2800 Lyngby, Denmark}

\date{\today}

\begin{abstract}
With the slow but constant progress in the coherent control of quantum systems, it is now possible to create large quantum superpositions. There has therefore been an increased interest in quantifying any claims of macroscopicity. We attempt here to motivate three criteria which we believe should enter in the assessment of macroscopic quantumness: The number of quantum fluctuation photons, the purity of the states, and the ease with which the branches making up the state can be distinguished.
\end{abstract}


\end{frontmatter}

\section{Introduction}

More than eighty years after its inception, quantum mechanics has become firmly established as a reliable model of the physical world. Even counter intuitive notions such as Schr\"{o}dinger's cat and wave-particle duality have trickled into the layman's vocabulary. Yet, to this day and even within the physics community, the coherent superposition of macroscopic objects still seems to intrigue more than does that of microscopic ones. One obvious reason for this is that, because of decoherence, large numbers of particles are difficult to shepherd into coherent ensembles. However, decoherence on its own does not account for the vagueness surrounding the macroscopicity buzzword \cite{Leggett1980}.

Several experiments, especially in solid state and atomic setups at cryogenic regimes, have exhibited quantized or coherent behaviour of macroscopic scales {\cite{Josephson1962, Anderson1995, Pereverzev1997}}. In quantum optics, coherent state superpositions, the so-called Schr\"{o}dinger cat states of light, have been generated and thoroughly studied for nearly a decade \cite{Ralph2003, Ourjoumtsev2006, NeergaardNielsen2006, Laghaout2013}. In view of these advances, the question of macroscopicity has shifted to a quantitative one: What observables make up the ``size'' of a quantum state? Several equally valid measures for this were proposed over the years \cite{Leggett2002, Bjork2004, Dur2002, Lee2011, Korsbakken2007, Sekatski2014a, Sekatski2014b, Wang2013, Volkoff2014, Frowis2012, Frowis2013}. Our purpose here is to give a bigger picture of the various prerequisites that macroscopicity entails. Indeed, the word ``macroscopicity'' has a dual etymology with ``macro'' meaning large, and ``scope'' alluding to an observer-dependent perspective. If, in addition, we talk of the \textit{quantum} macroscopicity of a system, we also expect that it exhibits quantum coherence. The criterion for quantum macroscopicity is thus three-fold: One should assert that a system is (1) large, (2) quantum, and (3) demonstrably composed of macroscopically distinct branches in at least some of its subsystems. 

The outline of this article is as follows. We begin by treating points (1) and (2) above in Sec. \ref{sec:ObjectiveMacroscopicity}, where we present a measure for the size of pure states which consists of the number $\mathcal{N}$ of fluctuation photons. Such a measure is objective in the sense that it is independent of the measurement process. We also give a brief reminder that the quantumness of a state is related to its purity and that the inclusion of purity in the macroscopicity measure is necessary, albeit non-trivial. Sec. \ref{sec:SubjectiveMacroscopicity} discusses  the observer's ability to distinguish mixed states. This is formalized with a distinguishability factor $\mathcal{D}$ which is then combined with $\mathcal{N}$ to produce what we shall refer to as the subjective---i.e., perceived---macroscopicity $\mathcal{M}$. By the same token, we emphasize that distinguishabiliy is fundamentally ill-defined for the branches of a coherent superposition.

\section{Objective macroscopicity}
\label{sec:ObjectiveMacroscopicity}

Our heuristic approach to macroscopicity begins with the phase space representation of physical states. Consider a classical state tracing a trajectory in phase space under some potential. It is represented by a geometrical point whose distance from the origin reflects how excited it is. In quantum mechanics, this point acquires a continuous pseudo-probability distribution---typically a Gaussian of finite width---of which it becomes the centroid. The canonical coordinates of the centroid are the same as in the classical picture \cite{Ehrenfest1927}; the first moment of the distribution is therefore unlikely to describe quantum properties. The second moment, on the other hand, arises from a coherent set of quantum fluctuations. It is these fluctuations, which amount to the quantum noise of the distribution, that are of interest. This means that, in effect, a coherent state of light is as macroscopic as the vacuum since the two only differ by their centroid's position. Even if the coherent state contains a larger number of photons, the effective number of them that contributes to quantum fluctuations is the same as that of the vacuum, namely zero. All the other photons can be considered as nothing more than a classical offset with no coherence content. 

From the motivation outlined above, we therefore propound that the macroscopicity of a quantum optical state is quantified by the mean number of photons minimized over all possible displacement operations. In other words, we define the macroscopicity as the number of photons associated with the fluctuations within a pure state $\ket{\psi}$,
\beqa
\mathcal{N}(\ket{\psi}) & = & \vin{\nhat_{\mbox{\scriptsize fluct.}}}_{\mbox{\scriptsize \ket{\psi}}} \nonumber\\
& = & \vin{\nhat} - \vin{\nhat_{\mbox{\scriptsize centroid}}} \nonumber\\
& = & \frac{1}{2}\tes{\mbox{var}(x) + \mbox{var}(p) - 1}, \label{eq:Nfluct}
\eeqa
with $\hat{n} = \frac{1}{2}(\hat{x}^2 + \hat{p}^2 - 1)$ and $\vin{\nhat_{\mbox{\scriptsize centroid}}} = \vin{\ahat}\vin{\adag} = \frac{1}{2}\tes{\vin{\hat{x}}^2 + \vin{\hat{p}}^2}$. The measure $\mathcal{N}$ is objective in the sense that it is expressed in physical units of fluctuation photons with no dependence on the measurement process. Further below, we shall also present a subjective---i.e., observer-dependent---version of it.

It is worth mentioning that, for pure states, the macroscopicity $\mathcal{N}$ coincides with that of Lee and Jeong \cite{Lee2011}, who arrived at their own measure from an entirely different motivation and which in turn matched some earlier results by D\"{u}r \textit{et al.} \cite{Dur2002}. We take this convergence of results as a strong indication that our heuristic described above is valid.

\begin{figure}[h]
  \centering\includegraphics[width=1\columnwidth]{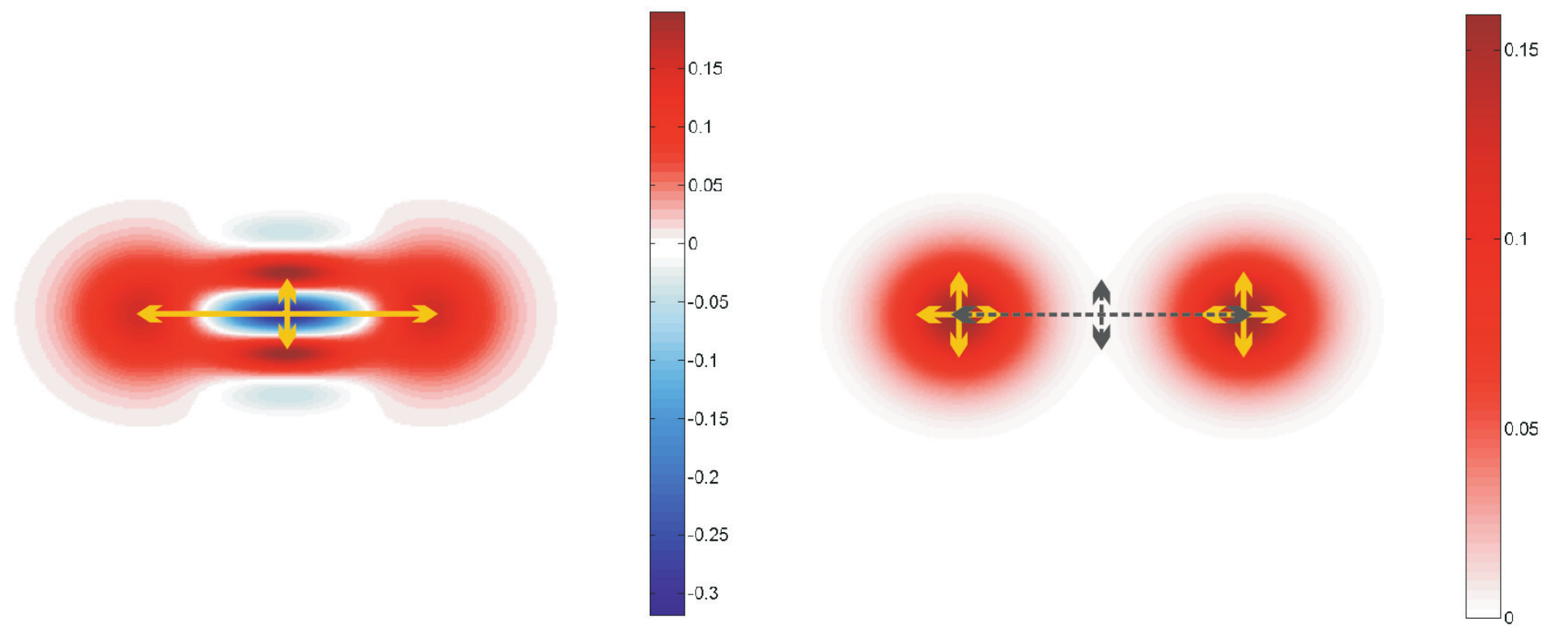}
  \caption{Wigner profiles of a coherent state superposition $\ket{\alpha}-\ket{{-}\alpha}$ (left) and a coherent state mixture $\ketbra{\alpha}{\alpha} + \ketbra{{-}\alpha}{{-}\alpha}$ (right) for $\alpha = 1.5$. The $x$- and $p$- variances are represented schematically by the orthogonal double arrows. The crucial difference between the superposition and the mixture is that the \textit{coherent} variance---i.e. the one that arises from quantum fluctuations, not classical statistics---is much smaller for the mixtures. Whereas it spans both lobes of the Wigner function for the superposition, its extent in the mixture is merely that of either coherent state $\ket{{\pm}\alpha}$. This ``genuinely quantum'' variance is represented by the yellow arrows and that is the one that should enter in the macroscopicity. Since both quantum and classical statistics get blended together in the Wigner function, a blind application of Eq. (\ref{eq:Nfluct}) will yield the wrong result for mixed states as it will mistake the overall variance (dotted grey) for a quantum variance.}
	\label{fig:CSSvariance}
\end{figure}

A warning is in order at this point regarding a critical caveat: The distributions of which we compute the second moment should be made up exclusively of coherent excitations. In other words, the derivation leading up to (\ref{eq:Nfluct}) only reveals genuine \textit{quantum fluctuation} photons provided the state under consideration has unit purity. Failing this, we lose track of whether the variance in the canonical coordinates is of quantum or classical origin since both distributions are blended indiscriminately into one and the same Wigner function. This is illustrated with the example of coherent state superpositions and mixtures in Fig. \ref{fig:CSSvariance}. The generalization of (\ref{eq:Nfluct}) which discerns the quantum second-moments from the classical ones is a non-trivial matter which we shall not attempt to tackle here. For the sake of simplicity, we shall therefore limit our discussion to pure states. (For a treatment of mixed states, we refer to the work of Lee and Jeong \cite{Lee2011}, who provide a general and intuitive strategy.)

\section{Subjective macroscopicity}
\label{sec:SubjectiveMacroscopicity}

We have so far presented the quantum size of an optical system as an objective measure that is observer independent. However, in the quantum optics community, the notion of macroscopicity is often associated with the subjective ability of an observer to distinguish the branches with a coarse-grained detector \cite{Korsbakken2007, Sekatski2014a, Sekatski2014b}. Using a ``classical'' detector such as the naked eye or a coarsed grained intensity detector one should be able to infer any one of the branches of the quantum state. The underpinning idea is that coarse-graining, for being insensitive to microscopic observables, can only discern macroscopically separated eigenvalues. 

It is therefore useful to define a subjective macroscopicity measure that involves the ability to distinguish pre-specified branches of a macroscopic quantum state. This not only contains information about how large a state is (the objective macroscopicity) but also information about how far apart its branches are from one another from the perspective of the observer. This follows the original spirit of Schr\"{o}dinger's thought experiment.   

\subsection{The notion of distinguishability}
\label{sec:Distinguishability}

We shall first elaborate on the notion of distinguishability in order to later incorporate it in a subjective measure of macroscopicity. The two concepts are often interlinked in the literature, as exemplified by the work of Korsbakken \textit{et al.} in \cite{Korsbakken2007}. A related strategy was followed by Sekatski \textit{et al.} \cite{Sekatski2014a, Sekatski2014b} who define macroscopicity by the ability to discriminate the branches using a classical-like intensity detector which cannot resolve photon numbers. Such a discrimination task of macroscopically distinct branches by using ideal homodyne detectors was considered in Ref \cite{Andersen2013}. Strongly motivated by these works (in particular by the work of Sekatski \textit{et al.}), we shall consider in this section the notion of distinguishability using a noisy detector and present various relevant examples.   

Recall Schr\"{o}dinger's original thought experiment: A macroscopic cat, which is entangled with an atomic qubit $\braces{\ket{\uparrow}, \ket{\downarrow}}$, is collapsed upon observation into either one of the two orthogonal states $\ket{\mbox{\small\sc dead}}$ or $\ket{\mbox{\small\sc alive}}$. For this simple two-level, two-mode system, the notion of distinguishability is essentially a measure of our confidence in being able to identify in a single-shot that the cat is either dead or alive. We shall express this confidence level as a number in the interval $\hak{0,1}$. The distinguishability will be affected by the choice of measurement basis and its concomitant resolution. Indeed, as observers, we are not directly comparing $\ket{\mbox{\small\sc dead}}$ with $\ket{\mbox{\small\sc alive}}$, but rather the probability distributions they imprint on our measurement devices. The problem of distinguishability therefore boils down to comparing probability distribution functions (PDFs). Several techniques for performing such comparisons are already known from statistics. In the next subsection, we shall opt for two of them, namely the Bhattacharyya coefficient and the Kolmogorov distance \cite{Bhattacharyya1943, Bhattacharyya1946, Fuchs1996, Fuchs1999}.

It is tacitly implied that the states to be distinguished are sampled on the measurement device from a classical ensemble, not a quantum superposition. In other words, looking at the second mode in 
\beq
\ket{\downarrow} \otimes \ket{\mbox{\small\sc dead}} +   \ket{\uparrow} \otimes \ket{\mbox{\small\sc alive}} \label{eq:Cat2mode}
\eeq 
upon the tracing out of the first mode, one essentially sees a classical mixture of $\braces{\ket{\mbox{\small\sc dead}}, \ket{\mbox{\small\sc alive}}}$. The comparison of the PDFs is then a perfectly legitimate thing to do since the PDF of a mixture is the mixture (read: average) of the PDFs. This is not the case if one is instead looking at a quantum superposition such as $c_1 \ket{\mbox{\small\sc dead}} + c_2 \ket{\mbox{\small\sc alive}}$, with $c_1, c_2 \in \mathbb{C}$. In this latter scenario, the PDF of the superposition will conceal a much more obscure mapping to the branches $\ket{\mbox{\small\sc dead}}$ or $\ket{\mbox{\small\sc alive}}$ depending on their relative phases and the orientation of the measurement basis.  For coherent superpositions, the very question of distinguishability thus becomes ill-defined, for similar reasons that which-path queries in a double-slit experiment are vain and counter-factual. 

With the above in mind, let us start by generalizing (\ref{eq:Cat2mode}) to encompass all pure states that are made up of $B \in \mathbb{N}$ distinct, pure, and normalized (though not necessarily orthogonal) branches $\braces{\ket{b_k}}$ such that
\beq
\ket{\psi} = \sum\limits_{k = 1}^{B} c_{k} \ket{q_k} \otimes \ket{b_k},
\label{eq:rhohat}
\eeq
where the coefficients $c_{k} \in \mathbb{C}$ satisfy the normalization condition and $\braces{\ket{q_k}}$ represent the entangling states in the first mode. For simplicity, these states are assumed to be distinct, pure, normalized and orthogonal. Collapses of the first mode on the orthogonal basis $\braces{\ket{q_k}}$ effectively produce the mixed state of branches 
\beq
\rhohat = \sum\limits_{k = 1}^{B} \abs{c_k}^2 \ketbra{b_k}{b_k}. \label{eq:MixedBranchesBis}
\eeq
in the second mode. The PDF of the state in (\ref{eq:MixedBranchesBis}) can then be sampled to infer (up to a certain confidence level) which branch $\braces{\ket{b_k}}$ was actually measured. The distinguishability in this case simply boils down to a straightforward comparison of PDFs. If, by contrast to the mixture in (\ref{eq:MixedBranchesBis}), we are interested in producing a coherent superposition of branches for which the macroscopicity is to be assessed, then the first mode needs to be collapsed on the diagonal basis. This will produce as many phase relationships within the superposition as there are diagonal vectors. This point will become clearer with the explicit examples of Sec. \ref{sec:Examples}.

\subsection{Measures of distinguishability}

Now that we have definitions for the states under consideration, let us formalize our notion of distinguishability. Let $\braces{\lambda}$ be the outcomes of the measurement operator $\Pihat_{\lambda}$ in the second mode. (E.g., $\lambda$ corresponds to photon counts for photon number resolving detectors or quadrature values for homodyne 
detection.) Ideally, $\Pihat_{\lambda} = \ketbra{\lambda}{\lambda}$. However, most detectors will have a finite resolution $\sigma$ that will lead to a ``blurring'' of the outcomes via the convolution with some kernel function $g_{\sigma}(\lambda')$,
\beq
\Pihat^{\sigma}_{\lambda} = \int\limits_{\lambda'} g_{\sigma}(\lambda-\lambda') \ketbra{\lambda'}{\lambda'} \, \mathrm{d}\lambda'. \label{eq:Pilambda}
\eeq
For simplicity, we shall model the non-ideal resolution $\sigma$ on the measurement spectrum by a Gaussian blur $g_{\sigma}(\lambda') = \frac{1}{\sigma\sqrt{2\pi}} e^{-\frac{\lambda'^2}{2\sigma^2}}$. (Should we refer to discrete measurement spectra, the integrals must be replaced by sums.) Upon the incidence of a normalized state $\rhohat$, one obtains a PDF over the range of outcomes
\beq
P(\lambda, \rhohat) = \mbox{Tr}\braces{\Pihat^{\sigma}_{\lambda} \rhohat}.
\eeq

One measure of distinguishability $\mathcal{D}$ between the branches that we propose here is
\beq
\mathcal{D}^{\mbox{\tiny BC}} = 1 - \sum\limits_{k=1}^{B} \abs{c_k}^2 \Omega(\ket{b_k}, \hat{\tilde{\rho}}_k),
\eeq
where $\Omega(\ket{b_k},  \hat{\tilde{\rho}}_k)$ is the Bhattacharyya coefficient between the PDF of branch $k$ and that of the weighted mixed set $\hat{\tilde{\rho}}_k$ of the remaining branches 
\beq
\hat{\tilde{\rho}}_k = \frac{\sum\limits_{l=1}^{B} (1-\delta_{l,k}) \abs{c_l}^2 \ketbra{b_l}{b_l}}{\mbox{Tr}\braces{\sum\limits_{l=1}^{B} (1-\delta_{l,k}) \abs{c_l}^2 \ketbra{b_l}{b_l}}}.
\label{eq:rhohattilde}
\eeq
The Bhattacharyya coefficient (BC) between the PDFs of two states $\rhohat_A$ and $\rhohat_B$ for a given measurement operator is given by
\beq
\Omega(\rhohat_A, \rhohat_B) = \int\limits_{\lambda} \sqrt{P(\lambda, \rhohat_A) \cdot P(\lambda, \rhohat_B)} \,\, \mathrm{d}\lambda
\eeq
and quantifies the amount of overlap between the two distributions in function space.

Another measure we also find pertinent to present is based on the Kolmogorov distance (KD). Unlike the BC, which is essentially an inner product that compares the PDFs in function space, the KD 
is closely related to the error probability (PE) in a standard binary decision problem through $\mathrm{KD} = 1 - 2\cdot \mathrm{PE}$. For two states with equal prior probabilities, it is defined (again for a specific measurement) as \cite{Fuchs1999} 
\beq
\mathrm{KD}(\rhohat_A, \rhohat_B) = \frac{1}{2} \int\limits_\lambda \abs{P(\lambda, \rhohat_A) - P(\lambda, \rhohat_B)} \,\mathrm{d}\lambda .
\eeq
Averaged over all the branches, it is given by
\beq
\mathcal{D}^{\mbox{\tiny KD}} = \sum\limits_{k=1}^{B}  \frac{\abs{c_k}^2}{2} \int\limits_\lambda \abs{P(\lambda, \ket{b_k}) -   P(\lambda, \hat{\tilde{\rho}}_k)} \, \mathrm{d}\lambda.
\eeq

\subsection{A measure for subjective macroscopicity}
\label{sec:SubjectiveMacroscopicityMeasure}

From the previous sections, we arrived at two measures, $\mathcal{N}$ and $\mathcal{D}$, which are  respectively independent and dependent on the measurement process. It may be convenient, however, to synthesize these two measures into a single number which we shall refer to as the subjective macroscopicity and which is simply the product
\beq
\mathcal{M} = \mathcal{N}\times\mathcal{D}.
\eeq
The underlying idea is that distinguishability, for being a number between 0 and 1, acts on $\mathcal{N}$ as a scaling factor to yield an ``effective'' number of fluctuation photons $\mathcal{M}$, i.e., the perceived macroscopicity of the system. Since $\mathcal{D}$, and therefore $\mathcal{M}$, are subjective measures, care should be taken to always specify the measurement device under consideration. Otherwise, the notion of macroscopicity cannot be used for comparative purposes between different pairs of quantum states and measurement devices.

\section{Examples}
\label{sec:Examples}

We shall now present the macroscopicity and distinguishability of a collection of two-branched states of the form 
\beq 
\ket{\psi} = \frac{\ket{\uparrow}\otimes\ket{b_1} + \ket{\downarrow}\otimes\ket{b_2}}{\sqrt{2}}.
\label{eq:twobranch}
\eeq

The distinguishability is assessed upon the projection of the first mode on the basis $\braces{\ket{\uparrow}, \ket{\downarrow}}$. That is, we try to discern the PDFs of $\ket{b_1}$ and $\ket{b_2}$. The macroscopicity, on the other hand is that of the superposition states $\ket{b_1} \pm \ket{b_2}$ and is obtained by projecting the first mode on the diagonal basis $\braces{\frac{1}{\sqrt{2}}\tes{\ket{\uparrow}\pm\ket{\downarrow}}}$. In practice, there will be errors in the projection on the qubit basis which need to be propagated into the computation of distinguishability. However, our analysis shall idealize the heralding operations on the first mode. 

\subsection{Coherent state superpositions}

One of the most archetypical examples of macroscopic superposition states is the coherent state superposition
$\ket{\psi^{\mbox{\tiny CSS}}_{\pm}} \propto \ket{\alpha} \pm \ket{{-}\alpha}$. This can be obtained from a qubit-coupled state \eqref{eq:twobranch} with branches

\beq
\ket{b_1} = \ket{\alpha}, \quad \ket{b_2} = \ket{{-}\alpha} .
\eeq

\begin{figure}[h]
  \centering\includegraphics[width=\columnwidth]{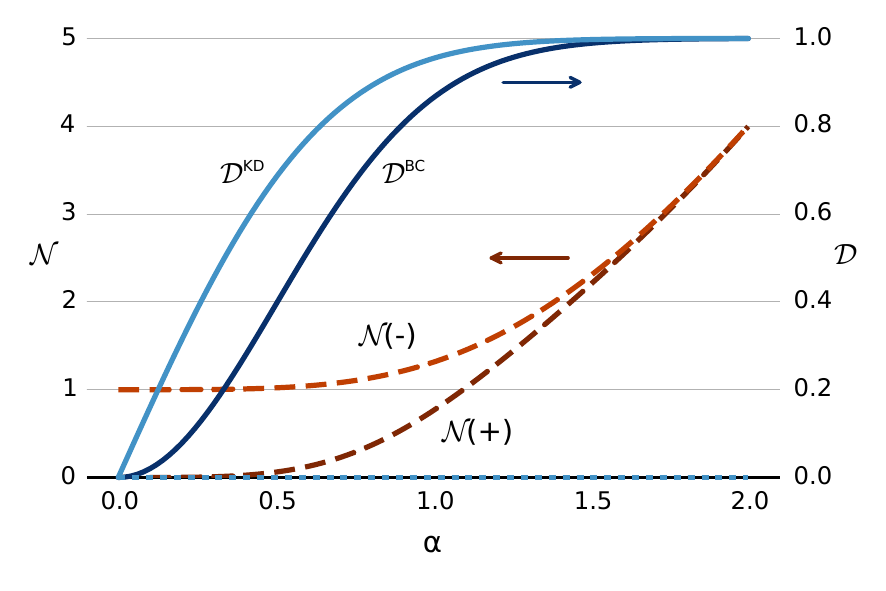}
  \caption{Objective macroscopicity (dashed) and distinguishability (solid) of $\ket{\psi^{\mbox{\tiny CSS}}_{\pm}}$ as a function of the amplitude $\alpha$. The constant dotted line lying along the abscissa represents the distinguishability using PNRDs.}
	\label{fig:CSS_Nfluct}
\end{figure}

In Fig. \ref{fig:CSS_Nfluct} we plot as a function of $\alpha$ the macroscopicity $\mathcal{N}$ of the superposition states $\ket{\psi^{\mbox{\tiny CSS}}_{\pm}}$, as well as the distinguishability of the two branches in terms of $\mathcal{D}^{\mbox{\tiny BC}}$ and $\mathcal{D}^{\mbox{\tiny KD}}$ for a homodyne measurement of the $x$ quadrature. 
The macroscopicities (number of fluctuation photons) are different for the two orthogonal superpositions but converging for large amplitudes. In that regime, the two coherent state branches are also clearly distinct as expected.
For smaller amplitudes the two $\mathcal{D}$ measures differ but follow the same trend.

If we were to use a photon number resolving detector (PNRD) instead, the two branches would be indistinguishable---at least without any additional displacement of the state.

Fig. \ref{fig:Summary} (a) plots the subjective macroscopicity $\mathcal{M}$, not as a function of the amplitude $\alpha$, but rather as a function of the \textit{total} number of photons $\vin{\hat{n}}$. This makes it possible to compare the macroscopic behavior of $\ket{\psi^{\mbox{\tiny CSS}}_{\pm}}$ with that of other photonic states which we present below and that depend on physical parameters other than $\alpha$. 

\subsection{Superposition of photon-subtracted vacua}

While true large-amplitude optical coherent state superpositions are extremely hard to prepare experimentally, photon-subtracted squeezed vacua provide an easier alternative with similar features. In Ref. \cite{Andersen2013} we proposed how superpositions of $m$- and $(m+1)$-photon subtracted states coupled to a microscopic mode could be created.
These states are perfectly distinguishable on an ideal PNRD, but for any realistic noise they would quickly blend together. On the other hand, consider the states in the rotated basis:

\begin{align}
\ket{b_1} &= \frac{N_m}{\sqrt{2}} \ahat^m \hat{S}(r) \ket{0} + \frac{N_{m+1}}{\sqrt{2}} \ahat^{m+1} \hat{S}(r) \ket{0}, \label{eq:SVS_b1}\\
\ket{b_2} &= \frac{N_m}{\sqrt{2}} \ahat^m \hat{S}(r) \ket{0} - \frac{N_{m+1}}{\sqrt{2}} \ahat^{m+1} \hat{S}(r) \ket{0}. \label{eq:SVS_b2}
\end{align}

\begin{figure}[h]
  \centering\includegraphics[width=\columnwidth]{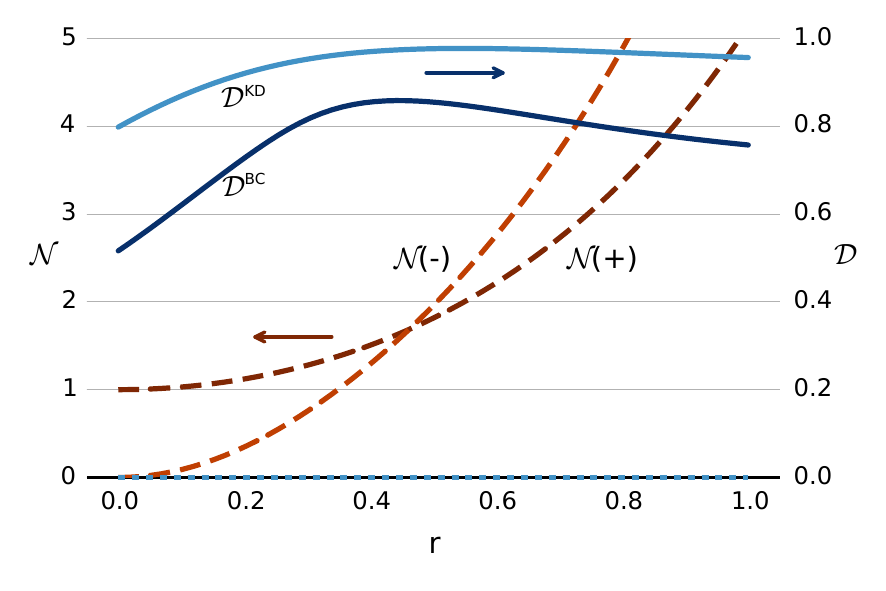}
  \caption{Objective macroscopicity (dashed) and distinguishability (solid) of $\ket{\psi^{\mbox{\tiny PSV}}_{\pm}}$ as a function of the squeezing parameter $r$.}
	\label{fig:SVS_Nfluct}
\end{figure}

$N_m$ normalizes the individual components. These branches are well separated in phase space and can therefore be distinguished even with a bad homodyne detector. 
As seen in Figure \ref{fig:SVS_Nfluct} (where we have taken $m=1$), this high distinguishability decreases slightly for very high squeezing levels due to the small side lobes of the quadrature distributions. The macroscopicity of $\ket{\psi^{\mbox{\tiny PSV}}_{\pm}} \propto \ket{b_1} \pm \ket{b_2}$ increases rapidly for increasing squeezing level. (Here again, measurements with a PNRD cannot produce any distinguishable PDFs, hence the dotted straight line along the abscissa.) The subjective macroscopicity of these states is illustrated in Fig. \ref{fig:Summary} (b).


\subsection{Displaced Fock state superpositions}

Our final, highly relevant example is the two-mode entangled single photon state $\ket{10} + \ket{01}$ which has undergone a displacement in the second mode. As before, the distinguishability is increased by considering the state in a rotated basis where the branches of \eqref{eq:twobranch} are

\begin{align}
\ket{b_1} &= \hat{D}(\alpha) \frac{ \ket{0} + \ket{1}}{\sqrt{2}}, \\
\ket{b_2} &= \hat{D}(\alpha) \frac{ \ket{0} - \ket{1}}{\sqrt{2}} .
\end{align}

\begin{figure}[h]
  \centering\includegraphics[width=\columnwidth]{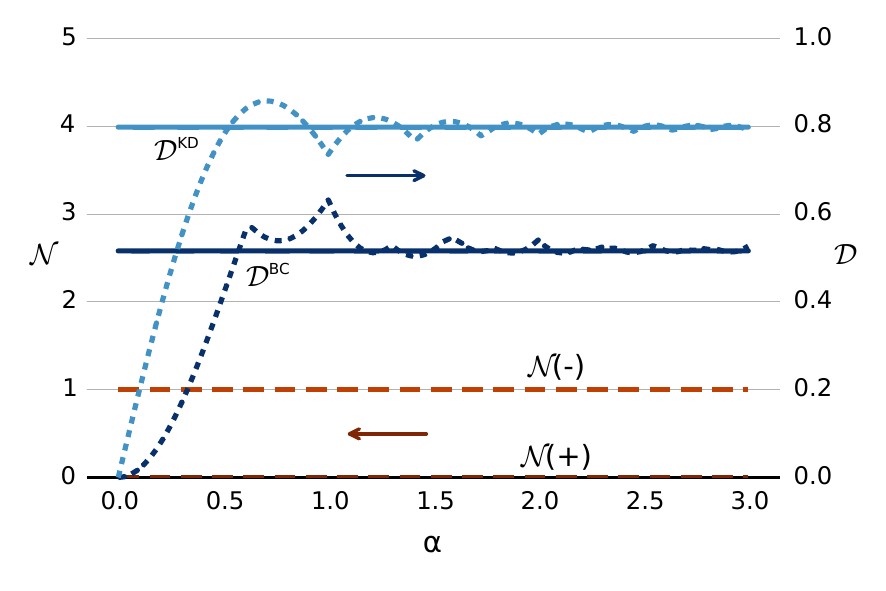}
  \caption{Objective macroscopicity (long dashes) and distinguishability of $\ket{\psi^{\mbox{\tiny DFS}}_{\pm}}$ as a function of the displacement amplitude $\alpha$. The distinguishabilities are given for a homodyne detector (solid) and a photon number resolving detector (short dashes).}
	\label{fig:FSS_Nfluct}
\end{figure}

This state was recently demonstrated as an example of a micro-macro entangled state, where the macroscopicity supposedly comes from the displacement of the state in phase space---a displacement that can be considerable \cite{Bruno2013, Lvovsky2013}. In light of our introductory discussion, however, such displacement does not correspond to an increase of the \emph{quantum} macroscopicity; it merely adds a classical amplitude to the state. With our definition of the macroscopicity \eqref{eq:Nfluct} we therefore see (Fig. \ref{fig:FSS_Nfluct}) a constant value of 0 (or 1) for $\ket{\psi^{\mbox{\tiny DFS}}_{\pm}} \propto \ket{b_1} \pm \ket{b_2}$. 

An interesting aspect of this state is that the branches are partially distinguishable with a homodyne as well as a PNR detector (solid and dashed blue curves, respectively). In fact, for large displacement amplitudes, $\mathcal{D}^{\mbox{\tiny BC}}$ and $\mathcal{D}^{\mbox{\tiny KD}}$ for the PNRD approaches the corresponding constant values for a homodyne detector. This is expected as an intensity detector effectively measures the amplitude quadrature for high intensities. Still, the distinguishability for any $\alpha$ is below that of the former class of superposition states (\ref{eq:SVS_b1}) and (\ref{eq:SVS_b2}) for any squeezing parameter $r$.


Note that the non-differentiable cusps in the distinguishability $\mathcal{D}^{\mbox{\tiny KD}}$ coincide with $\abs{\alpha} = \sqrt{m}$. This can be shown formally from the analytical expression of the states,
\beq
\ket{\psi_{\pm}^{\mbox{\tiny DFS}}} \propto  \frac{e^{-\frac{\abs{\alpha}^2}{2}}}{\sqrt{2}}  \sum\limits_{n=0}^{\infty} \frac{\alpha^n}{\sqrt{n!}} \tes{1 \mp \alpha^{*} \pm \frac{n}{\alpha}} \ket{n}, 
\eeq
which is obtainable from the generic expression for a displaced Fock state \cite{Kun2009}. A straightforward application of the Kolmogorov-based distinguishability, for example, yields
\beq
\mathcal{D}^{\mbox{\tiny KD}}(\alpha) = e^{-\abs{\alpha}^2} \mbox{Re}(\alpha) \sum\limits_{m=0}^{\infty} \frac{\abs{\alpha}^{2m-2}}{m!}\abs{m-\abs{\alpha}^2}
\eeq
where the term $\abs{m-\abs{\alpha}^2}$ contributes nothing to the cumulative sum whenever $\abs{\alpha}^2 \in \mathbb{N}^{+}$. A similar reasoning should lead to an explanation for the cusps in the Bhattacharyya-based distinguishability $\mathcal{D}^{\mbox{\tiny BC}}(\alpha)$. 

The subjective macroscopicity of these states is illustrated in Fig. \ref{fig:Summary} (c) where we see that it saturates almost immediately. Although larger displacements have little effect, a minimum of displacement is nonetheless crucial to achieve distinguishability (and hence macroscopicity) with the PNRD approach.

\subsection{Noisy detectors}

Realistic detectors will not be able to project sharply onto the eigenvalues of the measurement operator but will invariably be affected by noise. This will lead to a decreased distinguishability of quantum states. As mentioned in (\ref{eq:Pilambda}) above, detection noise can be modeled by a Gaussian distribution \cite{Sekatski2014a}. This model works better for a continuous-valued measurement like homodyne detection than for the discrete, positive eigenvalue spectrum of photon number detection, but it shall suffice for a preliminary understanding of the effect of detector noise.

\begin{figure}[h]
  \centering\includegraphics[width=\columnwidth]{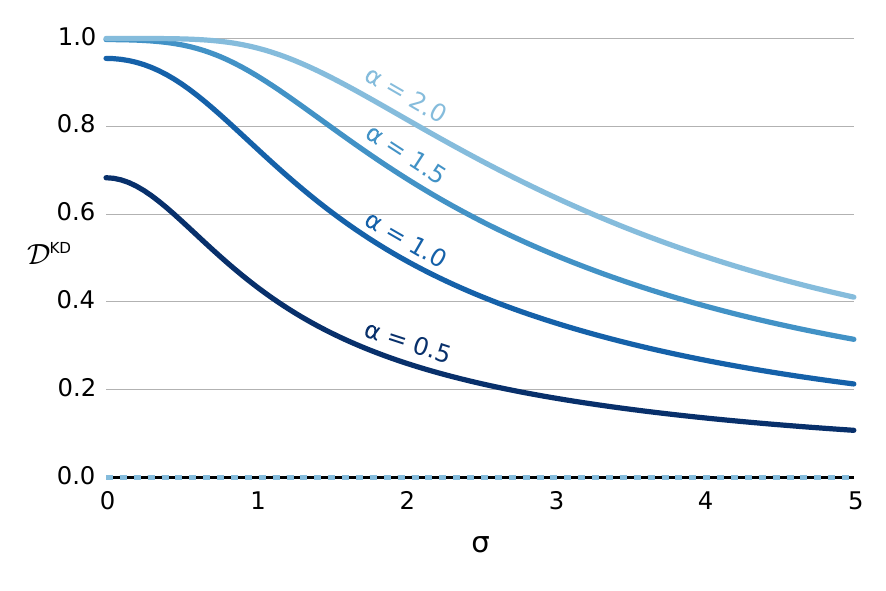}
  \caption{KD distinguishability of $\ket{\psi^{\mbox{\tiny CSS}}_{\pm}}$ as a function of detection resolution $\sigma$ for different coherent amplitudes $\alpha$.}
	\label{fig:CSS_KD}
\end{figure}

\begin{figure}[h]
  \centering\includegraphics[width=\columnwidth]{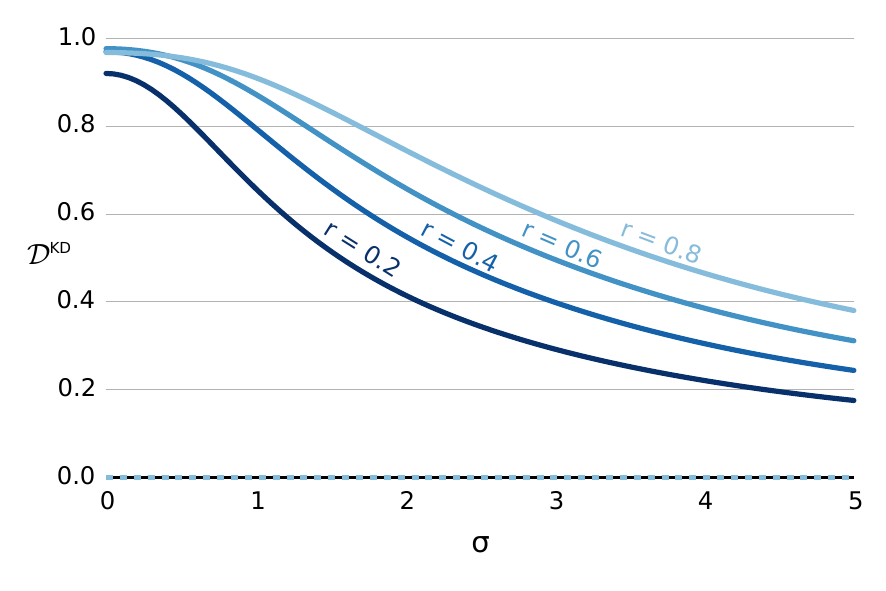}
  \caption{KD distinguishability of $\ket{\psi^{\mbox{\tiny PSV}}_{\pm}}$ as a function of detection resolution $\sigma$ for different squeezing parameters $r$.}
	\label{fig:SVS_KD}
\end{figure}

\begin{figure}[h]
  \centering\includegraphics[width=\columnwidth]{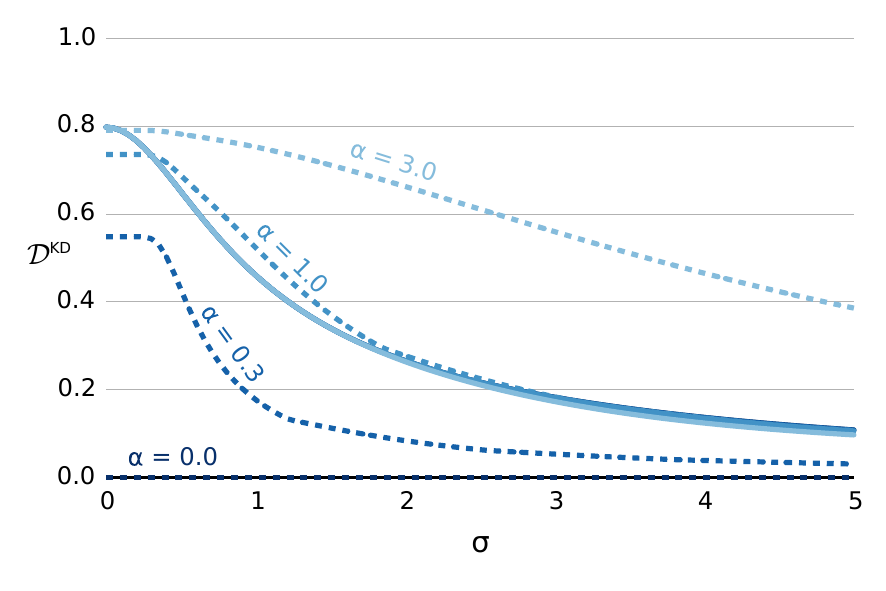}
  \caption{KD distinguishability of $\ket{\psi^{\mbox{\tiny DFS}}_{\pm}}$ as a function of detection resolution $\sigma$ for different  displacement amplitudes $\alpha$. Solid curves are for a homodyne detector, while the dashed curves are for a PNRD. Notice that the resolution $\sigma$ is not directly comparable between those two detectors.}
	\label{fig:FSS_KD}
\end{figure}

In Figs. \ref{fig:CSS_KD}--\ref{fig:FSS_KD} we plot the Kolmogorov distance $\mathcal{D}^{\mbox{\tiny KD}}$ as a function of the Gaussian detector blur $\sigma$ for the three classes of states with representative parameters. A common feature is that a higher amplitude, squeezing level, or displacement amplitude increases the states' tolerance towards detector noise. This can be interpreted as the states becoming more macroscopically distinguishable.

The displaced Fock state superposition (Fig. \ref{fig:FSS_KD}) deserves a special mention. For ideal detectors, the branches were essentially equally distinguishable whether detected in Fock or phase space. With a noisy detector, however, the distinguishability can be considerably higher when using a PNRD, at least for large displacements. In phase space, the displacement does not alter the probability distribution, so there is no gain.


\section{Summary}

We started with the premise that quantum fluctuations are the hallmark of quantumness. The more excitations of the field arise from these fluctuations, the larger the quantum size of the system. This lead us to define the quadrature variance of a state as the measure of its size. The physical unit of (objective) macroscopicity $\mathcal{N}$ is therefore the number of fluctuation excitations. Particular caution has to be exercised in order that the variance in question arises from the coherent---i.e., pure---components of the phase space distributions. This is a non-trivial task which we have avoided by focusing only on states of unit purity.

Furthermore, in the existing literature, the notion of macroscopicity often goes hand in hand with that of distinguishability \cite{Sekatski2014a, Sekatski2014b}. There could exist several different ways of looking at distinguishability $\mathcal{D}$ but we propose two which we believe are general enough and are based on the comparison of the PDFs of the branches making up a classical mixture. 

\begin{figure*}[h]
  \centering\includegraphics[width=1.95\columnwidth]{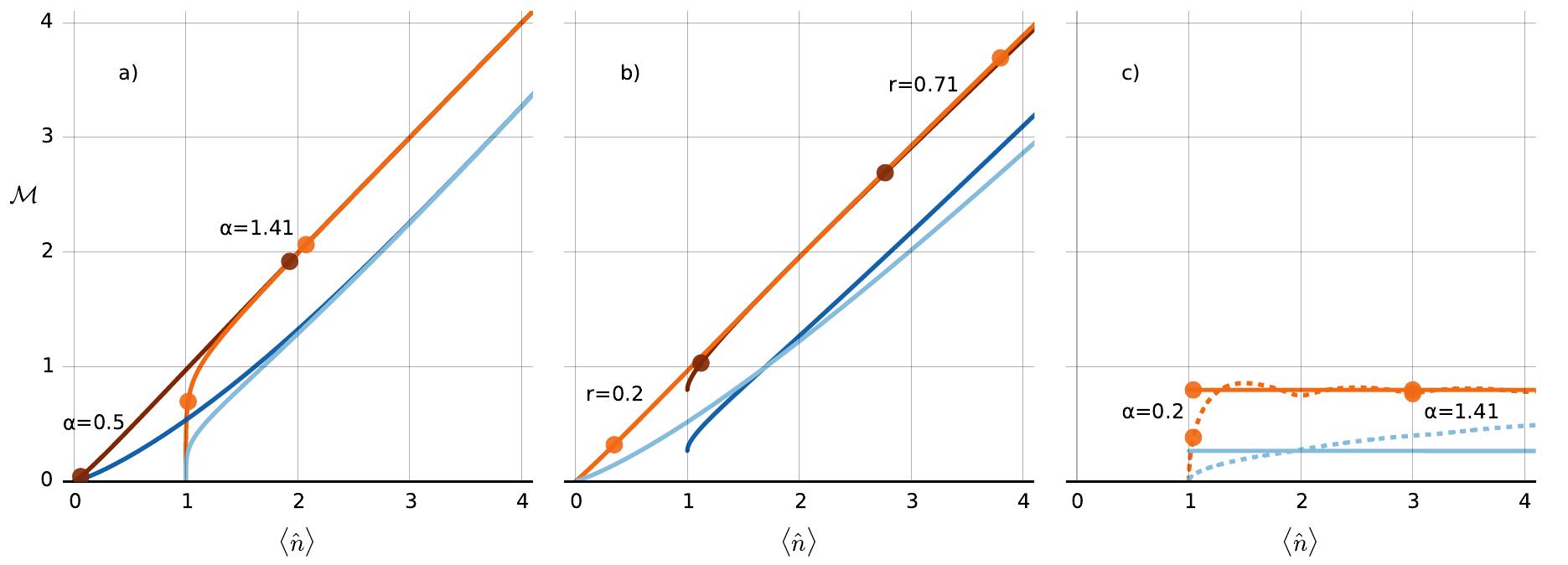}
  \caption{Subjective macroscopicity for (a) $\ket{\psi^{\mbox{\tiny CSS}}_{+}}$ (dark) and $\ket{\psi^{\mbox{\tiny CSS}}_{-}}$ (light), (b) $\ket{\psi^{\mbox{\tiny PSV}}_{+}}$ (dark) and $\ket{\psi^{\mbox{\tiny PSV}}_{-}}$ (light), and (c)  $\ket{\psi^{\mbox{\tiny DFS}}_{-}}$. The solid and dashed lines correspond to homodyne and photon-number resolving detection, respectively. Some fiducial markers indicate the physical parameters involved, namely coherent state amplitude, squeezing, and displacement, respectively. The brown/orange upper pairs of lines are obtained for an ideal measurement resolution $\sigma = 0$ whereas the blue lower pairs are for a non-ideal resolution of $\sigma = 2$. The experimentalist will be interested in reproducing this plot for resolutions $\sigma$ specific to the measurement devices at hand, keeping in mind that $\sigma$ is of a different nature for different measurement devices. (Note that the distinguishability used for $\mathcal{M}$ here is that of the Kolmogorov distance.)}
	\label{fig:Summary}
\end{figure*}

Finally, we argued for a subjective macroscopicity measure which combines $\mathcal{N}$ and $\mathcal{D}$. This is illustrated in Fig. \ref{fig:Summary} for various states with respect to the \textit{total} number of photons $\vin{\hat{n}}$. The reason for using the total number of photons in the abscissa is that it provides a common comparative reference for states which are otherwise parametrized by different units (e.g., squeezing, displacement). One can think of macroscopicity, in its largest sense, to be the region lying in the upper right corner, closest to---but not above---the diagonal line. As expected, the subjective macroscopicity is lowered away from the diagonal with a decrease in measurement resolution.

\section*{Acknowledgements}
{\noindent}This work was supported by the Danish Research Council. The authors  also acknowledge fruitful discussions with Clemens Sch\"{a}fermeier.


\section*{References}

\bibliography{mybibfile}

\end{document}